\documentclass[12pt]{article}

\usepackage{amsmath}
\usepackage{graphicx}
\usepackage{amsfonts}
\usepackage{amssymb}
\usepackage{latexsym}
\usepackage{color}
\usepackage{cite}
\input{colordvi.tex}

\renewcommand{\thefootnote}{\#\arabic{footnote}}

\begin{document}
\setcounter{footnote}{0}

\begin{titlepage}
\begin{flushright}
RESCEU-9/12
\end{flushright}
\begin{center}


\vskip .5in

{\Large \bf
Metric perturbation from inflationary magnetic field and
generic bound on inflation models
}

\vskip .45in

{\large
Teruaki Suyama$^1$
and 
Jun'ichi Yokoyama$^{1,2}$
}

\vskip .45in

{\em
$^1$
  Research Center for the Early Universe (RESCEU), Graduate School
  of Science,\\ The University of Tokyo, Tokyo 113-0033, Japan
  }\\
{\em
$^2$
  Institute for the Physics and Mathematics of the Universe (IPMU),\\
  The University of Tokyo, Kashiwa, Chiba, 277-8568, Japan
  }

\end{center}

\vskip .4in

\begin{abstract}
There is an observational indication of extragalactic magnetic fields.
No known astrophysical process can explain the origin of such large scale
magnetic fields,
which motivates us to look for their origin in primordial inflation.
By solving the linearized Einstein equations,
we study metric perturbations sourced by magnetic fields that are produced during inflation.
This leads to a simple but robust bound on the inflation models by requiring that
the induced metric perturbation should not exceed the observed value $10^{-5}$.
In case of the standard single field inflation model, the bound can be converted into
a lower bound on the Hubble parameter during inflation.
\end{abstract}
\end{titlepage}

\renewcommand{\thepage}{\arabic{page}}
\setcounter{page}{1}
\renewcommand{\thefootnote}{\#\arabic{footnote}}

\section{Introduction}
It has been well known that magnetic fields of $\sim \mu {\rm G}$ exist 
in galaxies (see for example, \cite{Kronberg:1993vk,Grasso:2000wj,Widrow:2002ud,Govoni:2004as}).
These fields are thought to be an outcome of the amplification 
of seed magnetic fields of unknown nature through the dynamo mechanism.
Magnetic fields are also known to exist in cluster of galaxies, too \cite{Kronberg:1993vk}.
Although such seed magnetic fields are expected to still reside in the intergalactic
medium and in voids, 
there had been no observational support of the seed magnetic fields until quite recently.
In 2010, observational data taken by Fermi and High Energy
Stereoscopic System (HESS) gamma-ray telescopes provided a strong 
support of the existence of extragalactic magnetic fields of at least 
$B \simeq 10^{-17}~{\rm G}$ on ${\rm Mpc}$ scales \cite{Neronov:1900zz,Taylor:2011bn,Vovk:2011aa}.
The lower bound on $B$ was derived from detection by HESS of ${\rm TeV}$ gamma rays 
coming from TeV blazars and non-detection by Fermi of ${\rm GeV}$ scale cascade emission, 
which is in contradiction with the assumption of zero magnetic field.

Explaining the origin of the extragalactic magnetic fields of 
$B \simeq 10^{-17}~{\rm G}$ on ${\rm Mpc}$ scales is challenging and 
remains to be addressed.
No promising astrophysical processes are known to generate 
the suggested amplitude of magnetic fields on such large length scales 
\footnote{For magnetic field generated from standard cosmological perturbations,
see \cite{Takahashi:2005nd,Ichiki:2006cd,Kobayashi:2007wd,Maeda:2008dv}.
For the effect of the primordial magnetic field on CMB temperature anisotropy and
its observational constraints, see \cite{Kosowsky:2004zh,Kahniashvili:2005xe,Kristiansen:2008tx,Kahniashvili:2010wm,Ichiki:2011ah,Shiraishi:2012rm}.}.
In light of this situation, 
the best alternative we can think of is to make use of inflation in the early Universe \cite{Guth:1980zm,Sato:1980yn,Starobinsky:1980te} which causally connects length scales
far beyond the Hubble radius through superluminal expansion
\cite{Ratra:1991bn,Gasperini:1995dh,Bamba:2003av,Bamba:2004cu,Bamba:2006ga,Martin:2007ue,Subramanian:2009fu,Kandus:2010nw,Barnaby:2012tk}.
It is well known that to realize magnetogenesis by inflation the action for the electromagnetic
field needs to be modified such that the action breaks conformal invariance.
The most-used Lagrangian for this purpose is a form $f^2(\phi) F^{\mu \nu} F_{\mu \nu}$,
where $\phi$ is some scalar field (e.~g.~ an inflaton or a dilaton) that varies during inflation.
It is not easy, however, to obtain a successful model, because typical models encounter
either breakdown of perturbation theory due to the strong coupling that occurs in the
earlier stage of inflation \cite{Demozzi:2009fu}, or generation of excessive electric
fields whose energy density may cause serious backreaction \cite{Ratra:1991bn,Bamba:2003av}.

The purpose of this paper is to discuss the latter effect caused by primordial magnetic
fields without respect to its specific generation mechanisms.
As already discussed in the literatures \cite{Kanno:2009ei,Finelli:2011cw,Byrnes:2011aa,Urban:2011bu,Demozzi:2012wh,Barnaby:2012tk},
if the energy density of the magnetic field exceeds the background inflaton
energy density, its backreaction onto the background dynamics becomes significant,
which generally destroys the homogeneity and isotropy of the Universe and 
puts a constraint on model building.
In addition to this, even if the backreaction effect is small and is not problematic
for background evolution,
there is another issue that one must take into account.
The energy-momentum tensor of the magnetic field also induces
metric perturbations and distorts the homogeneous and isotropic FLRW Universe \cite{Barrow:2006ch,Kojima:2009ms,Caldwell:2011ra}.
Degree of the distortion depends on the inflation models and also on the
electromagnetic models that achieve magnetogenesis.
But whatever be the model of inflationary magnetogenesis, 
the resultant amplitude of metric perturbation must not exceed the observed value $\sim 10^{-5}$ \cite{Komatsu:2010fb}.
Put it in another way, in addition to the standard constraints of explaining
the observed feature of primordial perturbations, 
we can derive an additional generic bound that must be imposed on any inflation model by 
analyzing metric perturbations arising from the inflationary magnetic fields,
which we work out in this paper.

To achieve this in quantitative manner, we study evolution of metric 
perturbations induced by magnetic field which is supposed to be generated during inflation.
We will treat the energy-momentum tensor of the magnetic field as first order perturbation and
solve the linearized Einstein equations on the FLRW Universe throughout the inflationary era
and the subsequent radiation dominated era.
This paper reports the results of such a calculation and derives a new constraint on
inflation models.
It is found that the slow-roll parameter that characterizes how slowly the Hubble
parameter changes during inflation needs to be bounded from below in order to
avoid large metric perturbation from magnetic field.
This is basically equivalent to the constraint $\delta \rho_{\inf} \gtrsim \delta \rho_B$
which tells that the fluctuation of inflaton energy density must be larger than
the energy density of the magnetic field.

\section{Metric perturbation from magnetic field}

In this section, we consider the evolution of the metric perturbation during inflation and 
subsequent radiation dominated Universe in the presence of magnetic field generated 
quantum mechanically by breaking the conformal invariance. 
Since magnetic field is assumed to be absent at the background level, 
we will treat its energy-momentum tensor as linear perturbation.
Perturbations of inflaton field and radiation/matter energy-momentum tensor are also
treated as linear perturbations.

Let us first write down the perturbed metric having only scalar-type perturbations in general gauge 
(for example, see \cite{Kodama:1985bj});
\begin{equation}
ds^2=-(1+2A) dt^2+2a^2 \partial_i B dx^i dt+a^2(t) \left[ (1+2\psi) \delta_{ij}+2 \partial_i \partial_j E \right] dx^i dx^j.
\end{equation}
Here $A,~B,~\psi$ and $E$ are metric perturbations.
We denote the perturbed energy-momentum tensor including both matter such as inflaton or radiation 
and magnetic field as 
\begin{equation}
\delta T^0_{~0}=-\delta \rho,~~~\delta T^0_{~i}=(\rho+P) a^2 \partial_i v,~~~\delta T^i_{~0}=-(\rho+P) \partial_i (v-B),~~~\delta T^i_{~j}=\delta^i_{~j} \delta P+\pi^i_{~j}, 
\end{equation}
where $v$ is defined by $u_i \equiv a^2 \partial_i v$ and
$\pi^i_{~i}=0$ is the anisotropic stress.
Since in many inflation models the anisotropic stress of matter is second order
in perturbation, we will assume that only the magnetic field yields the first order anisotropic stress.
If the time evolution of the magnetic field is solely due to the cosmological expansion,
$\pi^i_{~j}$ decays in proportional to $a^{-4}$ just in the same way as radiation.
On the other hand, if the electromagnetic action is modified, for example,
by having the electromagnetic field tensor couple to the inflaton like
\begin{equation}
S = -\frac{1}{4} \int d^4x~\sqrt{-g} f(\phi) F^{\mu \nu}F_{\mu \nu}, \label{ele-action}
\end{equation}
where $\phi$ is inflaton, then $\pi^i_{~j}$ can be made to decay much slower than $a^{-4}$.
In particular, in order to have scale invariant power spectrum of magnetic field
on cosmological scales, $\pi^i_{~j}$ needs to be almost constant during inflation.
This is indeed possible for some suitable choices of the functional form of $f(\phi)$.

For later convenience, we will write down the explicit transformation properties
of perturbation variables under the gauge transformation, 
$t \to {\bar t}=t-T,~x^i \to {\bar x}^i=x^i-\partial_i L$;
\begin{eqnarray}
&&{\bar A}=A+{\dot T},~~~{\bar B}=B+{\dot L}-\frac{T}{a^2},~~~{\bar \psi}=\psi+HT,~~~{\bar E}=E+L,~~~{\bar \sigma_g}=\sigma_g-\frac{T}{a^2}, \\
&&{\overline {\delta \rho}}=\delta \rho+{\dot \rho}T,~~~{\overline {\delta P}}=\delta P+{\dot P}T,~~~{\bar v}=v-\frac{T}{a^2},
\end{eqnarray}
where $\sigma_g \equiv B-{\dot E}$ denotes the shear of the four-velocity $u^\mu$.

The linearized Einstein equations in Fourier space are given by
\begin{eqnarray}
&&\frac{k^2}{a^2} \psi+Hk^2 \sigma_g-3H^2 A+3H {\dot \psi}=4\pi G \delta \rho, \\
&&HA-{\dot \psi}=-4\pi G(\rho+P) a^2 v, \\
&&-\frac{k^2}{3a^2} (A+\psi)+(3H^2+2{\dot H}) A-{\ddot \psi}-3H {\dot \psi}+H{\dot A}-\frac{k^2}{3} ( {\dot \sigma_g}+3H\sigma_g )=4\pi G\delta P, \\
&&\frac{k^2}{a^2} (A+\psi)+k^2 ( {\dot \sigma_g}+3H\sigma_g )=-8\pi G \Pi,
\end{eqnarray}
and $\Pi$ is defined by
\begin{equation}
\pi^i_{~j}=\left( \frac{1}{3} \delta_{ij}-\frac{k_i k_j}{k^2} \right) \Pi.
\end{equation}
The linearized conservation laws $\nabla_\mu T^\mu_{~\nu}=0$ are given by
\begin{eqnarray}
&&{\dot {\delta \rho}}+3H (\delta \rho+\delta P)+3(\rho+P) {\dot \psi}-(\rho+P) k^2 (v-\sigma_g)=0, \\
&&\partial_t \left[ (\rho+P)a^2v \right]+\delta P-\frac{2}{3} \Pi+3H(\rho+P)a^2v+(\rho+P)A=0.
\end{eqnarray}

To see how the curvature perturbation is affected by the presence of magnetic field,
let us consider the curvature perturbation on the uniform energy density hyper-surface
on which $\delta \rho=0$ is satisfied.
Denoting the curvature perturbation on this slice by $\zeta$,
we can derive its evolution equation by using both Einstein equations and
conservation laws given above. 
On superhorizon scales $k/(aH) \ll 1$, we find that the evolution equation reduces to
\begin{equation}
{\ddot \zeta}+3H {\dot \zeta}+\frac{1}{a^3}{\left( \frac{a^3 H}{\rho+P} \delta P_{\rm rel} \right)}^{\cdot}-\frac{8\pi G}{3} \Pi=0,
\end{equation}
where $\delta P_{\rm rel} \equiv \delta P_{\rm em}-\frac{\dot P}{\dot \rho} \delta \rho_{\rm em}$
($\delta \rho_{\rm em}$ and $\delta P_{\rm em}$ is the energy density and pressure 
perturbation for the electromagnetic field, respectively)
is the nonadiabatic pressure perturbation due to the relative entropy perturbation \cite{Kodama:1985bj,Malik:2002jb}
between the magnetic field and the component dominating the Universe \footnote{
Precisely speaking, intrinsic entropy perturbation for the dominant component $P_{\rm intr}$
will be correlated with $\delta P_{\rm rel}$.
But since the electromagnetic field is subdominant, 
it does not produce $P_{\rm intr}$ comparable to $\delta P_{\rm rel}$, or almost equal to $-P_{\rm rel}$.
Therefore, we do not need to consider the intrinsic entropy perturbation for the dominant component.
}.
Inhomogeneous solution of this differential equation is given by
\begin{equation}
\zeta (t)=-\int_{t_*}^t dt_1 \frac{H(t_1)}{\rho(t_1)+P(t_1)} \delta P_{\rm rel}(t_1)+\frac{8\pi G}{3} \int_{t_*}^t \frac{dt_1}{a^3 (t_1)} \int_{t_*}^{t_1} dt_2~a^3(t_2) \Pi (t_2), \label{zeta-mag}
\end{equation}
where we have imposed an initial condition that $\zeta (t_*)=0$.
In the actual situation, $t_*$ may be taken to be a horizon crossing time.
Equation (\ref{zeta-mag}) represents the magnetic field contribution to the
curvature perturbation.

Determining a precise value of the time integral in Eq.~(\ref{zeta-mag}) requires both model
specification and numerical computation, which is not what we want to do in this paper.
Instead, without referring to the specific inflation model,
we can estimate the order of magnitude on the basis of some generic properties of inflation. 
Let us first evaluate the first term containing $\delta P_{\rm rel}$ in the radiation 
dominated Universe achieved after reheating.
Noting that the electromagnetic tensor is traceless \footnote{
Traceless nature of the electromagnetic tensor persists even when the electromagnetic action 
is modified to a typical one given by Eq.~(\ref{ele-action}).
This may not be true for other modification of the Maxwell equations.
Eq.~(\ref{P-rho}) becomes an overestimate only if the effective equation of state parameter 
$w_m=P_m/\rho_m$ for the electromagnetic field becomes very close to $-1$, which is unlikely.
}, 
$\delta P_{\rm rel}$ during inflation can be approximated as
\begin{equation}
\delta P_{\rm rel} \simeq \frac{4}{3} \delta \rho_{\rm em}, \label{P-rho}
\end{equation}
where we have used an approximate relation ${\dot \rho} \simeq -{\dot P}$.
In the radiation dominated Universe, the energy density of magnetic field 
decays in the same way as that of radiation and we have $\delta P_{\rm rel}=0$.
Therefore, the first term yields a constant:
\begin{equation}
-\int_{t_*}^t dt_1 \frac{H(t_1)}{\rho(t_1)+P(t_1)} \delta P_{\rm rel}(t_1) \simeq -\frac{2 {\cal N}}{\epsilon} \frac{\delta \rho_{\rm em}}{\rho_{\rm inf}}, \label{integral-1st}
\end{equation}
where $\epsilon \equiv -{\dot H}/H^2$ is the slow-roll parameter,
$\rho_{\rm inf}$ is the energy density of the inflaton and ${\cal N}$
is the number of e-fold measured from the time when the mode of interest crossed the horizon
to the end of inflation.
Equation (\ref{integral-1st}) becomes exact if all the quantities are constant during inflation
but may deviate from the correct one if one of the quantities significantly changes during inflation.
While it is a good approximation to treat $\rho_{\rm inf}$ and $\epsilon$ as constants in many 
inflation models,
this does not necessarily hold for $\delta \rho_{\rm em}$.
In order to be conservative as possible as we can, let us replace Eq.~(\ref{integral-1st}) by
\begin{equation}
\bigg| \int_{t_*}^t dt_1 \frac{H(t_1)}{\rho(t_1)+P(t_1)} \delta P_{\rm rel}(t_1) \bigg| > \frac{1}{\epsilon} \bigg| \frac{\delta \rho_{\rm em}(t_{\rm end})}{\rho_{\rm inf}} \bigg|, \label{ineq-1st}
\end{equation}
where $t_{\rm end}$ is the time of inflation end.
A merit of this replacement is that, as we will see later, $| \delta \rho_{\rm em}(t_{\rm end})/ \rho_{\rm inf} |$
can be connected to the observed magnetic field strength and duration of
the dust-like Universe after inflation but before the reheating.
If $\delta \rho_{\rm em}$ is a growing function like $\delta \rho_{\rm em} \propto a^p~(p>0)$,
then left-hand side and right-hand side of the above inequality are almost the same.
If it is a decreasing function, then left-hand side becomes much larger than the right-hand side.
In either case, the left-hand side must not exceed the observed value of the primordial curvature perturbation
which is about $10^{-5}$.
If we further use another observational constraint that the curvature perturbation is almost Gaussian \cite{Komatsu:2010fb,Smidt:2010ra},
the bound can be made by a few orders of magnitude tighter.
This is because the curvature perturbation from Eq.~(\ref{ineq-1st}) is proportional to the magnetic
field squared and hence is highly non-Gaussian \cite{Caprini:2009vk,Seery:2008ms,Barnaby:2012tk}.
But since we do not know quantitatively how largely the non-Gaussianity from the magnetic field is 
allowed by observations, we decide to adopt the conservative bound that the curvature perturbation
from the magnetic field is smaller than $10^{-5}$.
This puts a non-trivial lower bound on $\epsilon$ as
\begin{equation}
\epsilon > 10^5 \times \bigg| \frac{\delta \rho_{\rm em}(t_{\rm end})}{\rho_{\rm inf}} \bigg|. \label{bound}
\end{equation}
As we will check later, inclusion of the second term of Eq.~(\ref{ineq-1st}) does not
alter this bound.
If one resorts to the inflationary magnetogenesis to explain the observed magnetic
fields on ${\rm Mpc}$ scales,
such inflationary model needs to satisfy this bound.

In case of the standard single field inflation model, the generic bound (\ref{bound}) can be rewritten
into the bound on combination of Hubble parameter during inflation $H_{\rm inf}$ and 
the so-called reheating parameter $R_{\rm rad}$ defined by,
\begin{equation}
R_{\rm rad}=\frac{a_{\rm end}}{a_{\rm reh}} {\left( \frac{\rho_{\rm end}}{\rho_{\rm reh}} \right)}^{1/4}=\exp \bigg[ \frac{\Delta N}{4}(-1+3w) \bigg],
\end{equation}
where $\Delta N$ is the number of $e$-fold between the end of inflation and the time of reheating
and $w=P/\rho$ is the equation of state parameter during that era.
This parameter was introduced and used in \cite{Martin:2006rs,Ringeval:2007am,Martin:2010kz,Demozzi:2012wh}
and roughly measures the degree of deviation from the radiation dominated expansion of and
duration of that era. 
Basic requirement that the inflation energy scale is less than ${(10^{-5}M_P)}^4$ and the reheating
occurs well before the Big-Bang Nucleosynthesis \cite{Kawasaki:1999na} restricts the allowed range of $R_{\rm rad}$ as
$-35< \log R_{\rm rad} <12$.
Using the standard formula for the curvature perturbation \cite{mukhanov} and observational data \cite{Komatsu:2010fb}:
\begin{equation}
{\cal P}_\zeta \simeq \frac{H_{\rm inf}^2}{8\pi^2 M_P^2 \epsilon} \simeq 2.4 \times 10^{-9},
\end{equation}
we can replace $\epsilon$ in Eq.~(\ref{bound}) by $H_{\rm inf}$.
Assuming no entropy is produced after reheating and using the reheating parameter,
Eq.~(\ref{bound}) can then be written as
\begin{equation}
\frac{H_{\rm inf}}{M_P} R_{\rm rad}^2 > 0.14 \times {\left( \frac{g_{*,{\rm reh}}}{g_{*,0}} \right)}^{1/6} {\left( \frac{\rho_{B,0}}{\rho_{\gamma,0}} \right)}^{1/2}
\simeq 9.0 \times 10^{-13} ~{\left( \frac{g_{*,{\rm reh}}}{200} \right)}^{1/6} \left( \frac{B}{10^{-17}~{\rm G}} \right), \label{bound2}
\end{equation}
where $g_*$ represents the relativistic degrees of freedom and $\rho_{\gamma,0}(\rho_{B,0})$
is the energy density of radiation(magnetic field) today.
This basically puts a lower bound on the energy scale of inflation.
In particular, if the inflation end is immediately followed by reheating or $w=\frac{1}{3}$,
in which case we have $R_{\rm rad}=1$,
the bound becomes
\begin{equation}
H_{\rm inf} > 2.2 \times 10^6~{\rm GeV} ~{\left( \frac{g_{*,{\rm reh}}}{200} \right)}^{1/6} \left( \frac{B}{10^{-17}~{\rm G}} \right).
\end{equation}
In the case the energy density of the magnetic field is diluted by $\xi <1$ compared to 
the dominant energy density due to the entropy injection after the reheating,
a factor $\xi^{-1/2}$ must be multiplied on the right-hand side of Eq.~(\ref{bound2}).
Therefore, if future observations detect tensor-to-scalar ratio, which uniquely fixes $H_{\rm inf}$,
and provide an decisive answer that the extragalactic magnetic fields of $B \simeq 10^{-17}~{\rm G}$
are of inflationary origin, 
we get a lower bound on $\xi$, namely, the maximum amount of entropy injection.

So far, we have focused on the first term on the right-hand side of Eq.~(\ref{zeta-mag}).
As for the second term, it does not dominate over the first term.
To see this, let us first notice that the integrand of the first term contains $\rho+P$ in
the denominator which enhances the integrand by $\epsilon^{-1}$ while the second term does not
have such an enhancement factor.
As a result, the contribution from inflationary era to the second term is suppressed 
by the slow-roll parameter $\epsilon$ compared to the first term. 
The second term receives an enhancement $\propto \log a$ in the radiation dominated era,
but this boosts the second term by at most $N_{\rm rad} < 100$,
the number of $e$-folds measured from the time of reheating.  
The second term can marginally reach the first term if $\epsilon$ is
as large as ${\cal O}(0.01)$ but becomes much smaller for smaller value
of $\epsilon$.
Therefore, apart from ${\cal O}(1)$ modification that may arise for inflation
models giving large $\epsilon$,
inclusion of the second term does not alter our bound (\ref{bound}).

Finally, before closing this section, we provide expressions for other perturbation
variables on the uniform density slice.
The Einstein equations and conservation laws can be recast into equations that give
$A$,~$\sigma_g$ and $v$ in terms of $\zeta$.
Using the solution (\ref{zeta-mag}) for $\zeta$, on super-horizon scales,
they are given by
\begin{eqnarray}
&&A=-\frac{\delta P_{\rm rel}}{\rho+P}, \\
&&k^2 \sigma_g = -\frac{8\pi G}{a^3} \int_{t_*}^t dt_1~a^3(t_1) \Pi (t_1), \label{uniform-sigma_g}\\
&&v =\frac{2}{3a^5 (\rho+P)} \int_{t_*}^t dt_1~a^3(t_1) \Pi (t_1).
\end{eqnarray}
These are at most the order of $\zeta$ or smaller than $\zeta$.

\section{Calculation in the Newtonian gauge}
In this section, we again consider the evolution of metric perturbations
in the Newtonian gauge which is widely used in the literature
including \cite{Bonvin:2011dt,Bonvin:2011dr} which obtained a physically incorrect result.
In the Newtonian gauge, the perturbed metric is given by
\begin{equation}
ds^2=-(1+2\Phi)dt^2+a^2(t) (1-2\Psi) \delta_{ij} dx^i dx^j.
\end{equation}
There is no gauge degree of freedom left in this coordinate system.
In this gauge, we find that the evolution equation of $\Psi$ can be written as
\begin{equation}
{\ddot \Psi}+\left( 4+\frac{3 {\dot P}}{\dot \rho} \right) H {\dot \Psi}+3H^2 \left( \frac{\dot P}{\dot \rho}-\frac{P}{\rho} \right) \Psi+c_s^2 \frac{k^2}{a^2} \Psi
=S,
\end{equation}
where $c_s^2$ is the sound speed of the dominant component
(e.~g.~, $c_s^2=\frac{\dot P}{\dot \rho}$ for the perfect fluid with barotropic equation of state
and $c_s^2=1$ for the canonical scalar field) and $S$ is defined by
\begin{equation}
S=\frac{8\pi G a^2 H}{\rho k^2} \bigg[ (\rho+P) {\left( \frac{\rho}{\rho+P} \right)}^\cdot \Pi+\rho \left( {\dot \Pi}+2H \Pi \right) \bigg]+4\pi G \left( \delta P_{\rm rel}-\frac{2}{3} \Pi \right). \label{def-S}
\end{equation}
Following the standard procedure (e.~g.~, the one given in \cite{mukhanov}), 
the inhomogeneous solution in the superhorizon regime ($k \ll aH$) is found to be
\begin{eqnarray}
\Psi = \frac{\sqrt{\rho(t)}}{a(t)} \bigg[ -&&\int_{t_*}^t dt_1 S(t_1) \frac{\sqrt{\rho(t_1)}}{\rho (t_1)+P(t_1)}
\int_{t_*}^{t_1} dt_2 ~\frac{a(t_2) \left( \rho (t_2)+P(t_2) \right)}{\rho (t_2)} \nonumber \\
&&+\int_{t_*}^t dt_1 ~\frac{a(t_1) \left( \rho (t_1)+P(t_1) \right)}{\rho (t_1)}
\int_{t_*}^t dt_1 S(t_1) \frac{\sqrt{\rho(t_1)}}{\rho (t_1)+P(t_1)} \bigg]. \label{sol-Psi}
\end{eqnarray}
Substituting Eq.~(\ref{def-S}) and performing integration by parts, 
this can be further recast into another form which makes its time dependence more transparent:
\begin{eqnarray}
\Psi = &&\frac{8\pi GH(t)}{a(t) k^2} \bigg[ \int_{t_*}^t dt_1~a^3(t_1) \Pi (t_1) -\frac{a^2 (t_*) \rho (t_*) \Pi (t_*)}{\rho (t_*)+P(t_*)}\int_{t_*}^t dt_1 ~\frac{a(t_1) \left( \rho (t_1)+P(t_1) \right)}{\rho (t_1)}\bigg] \nonumber \\
&&+\frac{H(t)}{a(t)} \int_{t_*}^t dt_1~\frac{H(t_1)}{\rho(t_1)+P(t_1)} \left( \delta P_{\rm rel}(t_1)-\frac{2}{3} \Pi(t_1) \right) \left( \frac{a(t)}{H(t)}-\int_{t_1}^t dt_2~a(t_2) \right). \label{final-Psi}
\end{eqnarray}
This is our final form of $\Psi$ induced by magnetic field.

Now, let us first consider the behavior of this solution during inflationary era. 
Just for simplicity, we assume the single field inflation model and treat $\Pi,~\rho,~P$ 
and hence $\epsilon$ too as very slowly changing variables.
Then, the first term of Eq.~(\ref{final-Psi}) gives the dominant contribution.
Neglecting the other terms, Eq.~(\ref{final-Psi}) during inflation becomes
\begin{equation}
\Psi \simeq \frac{8\pi G a^2}{3k^2} \Pi ={\left( \frac{aH}{k} \right)}^2 \frac{\Pi}{\rho}. \label{Psi-inf}
\end{equation}
This shows that the curvature perturbation in the Newtonian gauge grows very 
rapidly in proportion to $\propto a^2$.
Since $\Pi$ has the same order of magnitude as the energy density of the magnetic field,
$\Psi$ is suppressed by a small number, which is the fraction of the magnetic field 
energy density to the total one. 
But it is also enhanced by another big number which is square of 
the ratio of the wavelength of the mode to the horizon radius.
As a demonstration, if we approximate $\Pi/\rho$ at the end of inflation by
the ratio of magnetic field energy density to the CMB energy density today
and assume modes corresponding to $1~{\rm Mpc}$ today have expanded 50 $e$-folds
after the horizon exit, $\Psi$ at the end of inflation becomes as large as 
$3\times 10^{20}$ for $B = 10^{-17}~{\rm G}$ today.
This rough estimate is enough to see that $\Psi$ induced by inflationary magnetic field 
corresponding to $10^{-17}~{\rm G}$ today becomes quite large at the end of inflation.

Let us next consider the time evolution of $\Psi$ in the radiation dominated era,
assuming that reheating completes instantly just after inflation.
Using the conservation law $\Pi a^4 ={\rm const.}$ in the radiation dominated Universe,
we have
\begin{equation}
\int_{t_*}^t dt_1~a^3 (t_1) \Pi (t_1) \simeq \frac{\Pi (t) a^3 (t)}{H(t)}.
\end{equation}
Therefore, the contribution from the first term to $\Psi$ in Eq.~(\ref{final-Psi}) becomes
\begin{equation}
\Psi_{\rm first} \simeq 3 {\left( \frac{aH}{k} \right)}^2 \frac{\Pi}{\rho}. \label{Psi-first}
\end{equation}
Apart from the numerical factor of ${\cal O}(1)$,
this takes the same form as that in the inflationary era (see Eq.~(\ref{Psi-inf})).
However, their time dependence is quite different.
Indeed, in the radiation dominated Universe, $\Psi_{\rm first}$ decays in proportion to $a^{-2}$,
which significantly reduces the amplitude of $\Psi_{\rm first}$ that was quite large
at the time of reheating.
In particular, at the time of horizon reentry, $\Psi_{\rm first}$ becomes as small as $\Pi /\rho$.
For the magnetic field corresponding to $B=10^{-17}~{\rm G}$ today, 
this becomes $\Pi /\rho \sim 10^{-23}$, which is negligibly small.
Therefore, contributions from the remaining terms may eventually dominate over $\Psi$ at late time,
especially at the time of horizon reentry.
Given that the anisotropic stress and the energy density of magnetic field are
of the same order of magnitude,
we can show that magnitude of $\Psi$ at the time of horizon reentry is roughly given by
Eq.~(\ref{integral-1st}).
Since what we observe is $\Psi$ after the horizon reentry, the observational bound
$\Psi \lesssim 10^{-5}$ must be imposed at the time of horizon reentry,
which yields the same bound as Eq.~(\ref{bound}).
Therefore, as it should be, we have confirmed that consideration in the Newtonian gauge
yields the same bound as the one in the uniform density slice.

The feature $\Psi \propto \Pi a^2/k^2$ during inflation was also found in \cite{Bonvin:2011dt,Bonvin:2011dr}.
It was then claimed in \cite{Bonvin:2011dt,Bonvin:2011dr} that the large amplitude of $\Psi$ at the
end of inflation generically enters a constant mode in the radiation dominated era,
making the Universe eventually highly inhomogeneous and spoiling the success of standard
cosmology based on the FLRW metric.
This observation was derived by using matching conditions formulated in \cite{Deruelle:1995kd}
under an assumption of instant transition from the end of inflation to reheating.
However, as we have shown above, we do not find such contamination to the constant mode.
Instead, we observe that the growing magnetic mode $\Psi \propto \Pi a^2/k^2$,
which was significantly enhanced by the end of inflation, 
is taken over only by the decaying mode in the radiation dominated era.
Therefore, the breaking of the FLRW metric in the radiation dominated era is not a 
generic consequence of the inflationary magnetogenesis.

We can also rederive the time evolutionary behavior of $\Psi$ by connecting it with perturbations
evaluated in the uniform density slice using the gauge transformation,
which enables us to understand intuitively the origin of the rapid growth 
$\Psi \propto a^2$ during inflation. 
Noting that the shear $\sigma_g$ is identically zero in the Newtonian gauge,
time translation $T$ connecting the uniform density slice and the Newtonian gauge
is uniquely determined by $\sigma_g$ evaluated in the uniform density slice (see Eq.~(\ref{uniform-sigma_g}));
\begin{equation}
T=-\frac{8\pi G}{ak^2} \int_{t_*}^t dt_1~a^3(t_1) \Pi (t_1). \label{time-translation}
\end{equation}
Assuming $\Pi$ is almost constant during inflation, we see that $T$ grows in proportion to $a^2$.
This means that $t={\rm const.}$ hypersurface in the uniform density slice and that
in the Newtonian gauge deviate more and more with time.
Using the gauge transformation rule for $\psi$, the curvature perturbation $\Psi$ is then given by
\begin{equation}
\Psi=-\zeta+\frac{8\pi G \Pi}{3} \frac{a^2}{k^2}.
\end{equation}
The second term grows rapidly in time and eventually dominates over $\Psi$.
The Universe becomes apparently inhomogeneous with time in the Newtonian gauge, but this is
simply because the time translation (\ref{time-translation}) becomes very large.
From a geometric point of view, the rapid growth of $T$ reflects mismatch between spacetime having anisotropic stress 
even on large scales and the Newtonian gauge in which the spacial metric looks isotropic.
Therefore, the rapid growth of inhomogeneity does not occur on the slicing such as uniform density slicing
where the shear $\sigma_g$ does not vanish.

In the radiation dominated era, calculation of Eq.~(\ref{time-translation}) shows that
$T$ becomes constant in time.
Since $H$ decays in proportion to $a^{-2}$, the gauge transformation rule for the curvature
perturbation tells us that difference between $\Psi$ and $\zeta$ becomes smaller and smaller
with time.
In particular, $\Psi$ and $\zeta$ become of the same order of magnitude at the time of horizon reentry.

\section{Summary}

We have studied time evolution of metric perturbation induced by magnetic field
that is supposed to be produced during inflation.
Treating the energy-momentum tensor of the magnetic field as first order
perturbation, we solved the linearized Einstein equations on FLRW background
spacetime.
The resultant metric perturbation is at least of the order of 
$\delta \rho_B/(\epsilon \rho_{\rm inf})$, where $\delta \rho_B$ is the typical
amplitude of the magnetic field at the end of inflation,
$\epsilon$ is the slow-roll parameter and $\rho_{\rm inf}$ is the total energy 
density during inflation.
This perturbation must not exceed the observed amplitude $\sim 10^{-5}$,
from which we could obtain a generic bound on inflation models,
namely, inequalities (\ref{bound}) and (\ref{bound2}).
Any inflationary model that achieves magnetogenesis must satisfy these constraints.

We also performed perturbation analysis in the Newtonian gauge in which perturbations 
were found to rapidly grow during inflation on super-horizon scales.
This weird behavior is simply an artifact of mismatch of the Newtonian gauge 
which uses isotropic coordinate and the anisotropic stress of magnetic field that 
persists even on superhorizon scales.
Thus we do not have to worry about the breakdown of the FLRW background contrary
to the claim of \cite{Bonvin:2011dt,Bonvin:2011dr}.
The apparently enhanced perturbation by the end of inflation in this gauge starts to attenuate 
in the subsequent radiation dominate Universe.
The requirement that the perturbation must be smaller than $\sim 10^{-5}$
at the time of horizon reentry leads to the same bound as above.\\

\noindent {\bf Acknowledgments:} 
This work was supported by a Grant-in-Aid for JSPS Fellows
No.~1008477(TS), JSPS  Grant-in-Aid for Scientific Research
No.\ 23340058 (JY), and the Grant-in-Aid for
Scientific Research on Innovative Areas No.\ 21111006 (JY).

\end{document}